\documentclass[twocolumn,prl,aps,showpacs,superscriptaddress,notitlepage]{revtex4}
\usepackage[latin9]{inputenc}
\usepackage{amsmath}
\usepackage{amssymb}
\usepackage{graphicx}
\usepackage{epsfig}
\usepackage{amsfonts}
\usepackage{bbm}
\usepackage{psfrag}
\usepackage{rotating}
\usepackage{graphics}
\usepackage{txfonts}

\makeatletter

\usepackage{dcolumn}
\usepackage{bm}

\makeatother

\begin{document}

\title{Strong micro-macro entanglement from a weak cross-Kerr nonlinearity}
\author{Tian Wang}
\address{Institute for Quantum Science and Technology and Department of Physics
and Astronomy, University of Calgary, Calgary T2N 1N4, Alberta, Canada}
\author{Hon Wai Lau}
\address{Institute for Quantum Science and Technology and Department of Physics
and Astronomy, University of Calgary, Calgary T2N 1N4, Alberta, Canada}
\author{Hamidreza Kaviani}
\address{Institute for Quantum Science and Technology and Department of Physics
and Astronomy, University of Calgary, Calgary T2N 1N4, Alberta, Canada}
\author{Roohollah Ghobadi}
\address{Institute for Quantum Science and Technology and Department of Physics
and Astronomy, University of Calgary, Calgary T2N 1N4, Alberta, Canada}
\address{Institute of Atomic and Subatomic Physics, TU Wien, Stadionallee 2, 1020 Wien, Austria}
\author{Christoph Simon}
\address{Institute for Quantum Science and Technology and Department of Physics
and Astronomy, University of Calgary, Calgary T2N 1N4, Alberta, Canada}

\begin{abstract}
We study the entanglement generated by a weak cross-Kerr nonlinearity between two initial coherent states, one of which has an amplitude close to the single-photon level, while the other one is macroscopic. We show that strong micro-macro entanglement is possible for weak phase shifts by choosing the amplitude of the macroscopic beam sufficiently large. We analyze the effects of loss and discuss possible experimental demonstrations of the micro-macro entanglement based on homodyne tomography and on a new entanglement witness.
\end{abstract}
\maketitle

 Kerr nonlinearities (or photon-photon interactions) in appropriate media are both interesting from a fundamental quantum optics point of view and attractive for photonic quantum information processing. Strong Kerr nonlinearities in the optical domain are being studied in various systems including cavity QED \cite{cavity} and Rydberg states \cite{rydberg}. However, weak nonlinearities, such as those that can be induced in atomic ensembles by the AC Stark shift, have recently also received a lot of attention, in particular for quantum non-demolition (QND) detection \cite{QND}, but also for other tasks in quantum information processing \cite{weak-QIP}. The protocols considered for weak nonlinearities typically involve light at the single-photon level interacting with one or more strong coherent beams \cite{HePRA}. This is possible because, in contrast to most schemes for strong photon-photon interactions, weak nonlinearities are usually not  saturated by light at the single photon level, allowing the use of macroscopic input beams. There has recently been significant experimental progress on weak cross-Kerr nonlinearities \cite{weak-experiments}. In particular, Ref. \cite{GaetaNP} demonstrated a cross-Kerr phase shift of 0.3 mrad per photon using an atomic vapor in a hollow-core photonic bandgap fiber.

The creation of quantum effects such as superpositions and entanglement involving mesoscopic or macroscopic systems is currently being pursued in a number of areas including atoms \cite{atoms,oberthaler}, molecules \cite{molecules}, superconducting circuits \cite{supercond}, optomechanics \cite{verhagen,lehnert}, and purely optical systems \cite{light}. In particular, micro-macro entanglement of light was generated by displacing single-photon entangled states in phase space \cite{lvovskybruno}. The entanglement in the latter experiments was relatively weak. Inspired by the recent progress on weak Kerr nonlinearities and on macroscopic quantum states of light, we here study the entanglement generated by a weak cross-Kerr nonlinearity between two coherent beams, one of which is close to the single-photon level, while the other one is macroscopic. As mentioned above, this regime is particularly relevant for ongoing experimental research programs in quantum information processing. We show that strong micro-macro entanglement  is possible for weak per-photon phase shifts by choosing the amplitude of the macroscopic beam sufficiently large. In the absence of loss, increasing the weaker field increases the entanglement. This behavior changes in the presence of loss, where the entanglement starts to decrease again if the weaker amplitude is increased too much. We also discuss potential experimental demonstrations, based on homodyne tomography and on an entanglement witness that we introduce below.

\begin{figure}
\begin{centering}
\includegraphics[width=0.8\columnwidth]{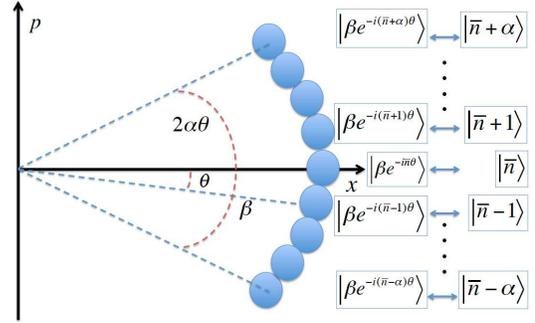}
\par\end{centering}
\caption{Pictorial representation of the pure entangled state Eq. (\ref{eq:pure}) created by a cross-Kerr nonlinearity starting from two initial coherent states $|\alpha\rangle |\beta\rangle$ in the absence of loss. Fock states $|\bar{n}+k\rangle$ (where $\bar{n}=\alpha^2$) for the first mode are correlated with coherent states with phases proportional to $(\bar{n}+k)\theta$ for the second mode, where $\theta$ is the weak cross-Kerr phase shift. The same correlation also holds when the roles of the modes are exchanged, see Eq. (1). Increasing $\theta$ causes a larger separation between neighboring coherent states and thus results in larger entanglement until neighboring coherent states are well separated in phase space, at which point the entanglement saturates. The total entanglement is bounded by the number of significant (well-separated) terms in the expansion Eq. (\ref{eq:pure}), which is of order $2 \alpha$ (assuming $\alpha \leq \beta$).}
\end{figure}

We now describe our results in more detail. We first discuss the lossless case. The Hamiltonian for a cross-Kerr medium is $H=\chi \hat{n}_1 \hat{n}_2$. A weak nonlinearity is described by a time evolution operator $U=e^{-i t H}=e^{-i\theta\hat{n}_1 \hat{n}_2}$, where $\theta=\chi t \ll 1$. When $U$ acts on two initial coherent states $|\alpha\rangle_{1}\ensuremath{|\beta\rangle_{2}}$ one obtains the state
\begin{equation}
|\psi\rangle=\sum_{n=0}^{\infty}{\displaystyle e^{-\frac{\alpha^{2}}{2}}\frac{\alpha^{n}}{\sqrt{n!}}|n\rangle_{1}|\beta e^{-in\theta}\rangle_{2}}=\sum_{n=0}^{\infty}{\displaystyle e^{-\frac{\beta^{2}}{2}}\frac{\beta^{n}}{\sqrt{n!}}|\alpha e^{-in\theta}\rangle_{1}|n\rangle_{2}},\label{eq:pure}
\end{equation}
where we have taken $\alpha$ and $\beta$ real for now for simplicity. One sees that the photon number in mode 1 is perfectly correlated with the phase in mode 2 and vice versa, see also Fig. 1. These number-phase correlations are the basis for our entanglement witness derived below.

We now analyze the amount of entanglement in the state of Eq. (1). We focus on the micro-macro regime $\beta \gg \alpha$ and on the first expansion in Eq. (1), which is in terms of photon number states for mode 1 and coherent states with different phases for mode 2. The overlap of two coherent states with different phases is
$|\langle\beta e^{-in\theta}|\beta e^{-im\theta}\rangle|^{2}=e^{-2\beta{}^{2}(1-cos((m-n)\theta))}\approx e^{-(\beta\theta(m-n))^{2}}$. As a consequence, for $\beta\theta \gtrsim 1$, subsequent coherent states in the expansion are close to being orthogonal, see also Fig. 1. The state is clearly non-Gaussian in this regime, since the peaks corresponding to different coherent states in Fig. 1 are well-separated. Even for very small values of $\theta$ there is always a value of $\beta$ large enough to achieve this condition. Note that the same condition is also required for quantum non-demolition detection \cite{QND}. Increasing $\beta$ (or $\theta$) further beyond this point does not increase the amount of entanglement significantly. The total amount of entanglement is then determined by the number of significant bi-orthogonal terms in the expansion, which is governed by the Poisson distribution for mode 1 (still assuming that $\alpha \ll \beta$). This number is of order  $2\alpha$, see also Fig. 1. One can thus increase the amount of entanglement at will by increasing $\alpha$.

However, our discussion so far was for the ideal case. We now analyze how the entanglement is affected by photon loss. We focus on loss after the state has been created (such as detection loss), since loss before the nonlinear operation only changes the amplitude of the coherent states, and loss during the operation can be made quite small in practice \cite{GaetaNP}.

Losses can be modeled by beam splitter operations between modes 1 and 2 and vacuum modes 3 and 4 respectively, of the form $|\alpha\rangle_1|0\rangle_3\rightarrow|t_1\alpha\rangle_1|r_1\alpha\rangle_3$ with $t_1^2+r_1^2=1$ etc. From the first identity in Eq. (1), one can e.g. first treat the loss in mode 2 in this way. The key step is to expand the coherent states in modes
2 and 4 and recombine the relevant terms in order to write mode 1 in terms of coherent
states. Then we can treat the loss in mode 1 simply by performing a beam splitter operation with mode 3. This gives the four-mode state
\begin{eqnarray}
\sum_{m,k=0}^{\infty}e^{-\frac{\beta{}^{2}}{2}}\frac{(t_{2}\beta)^{m}}{\sqrt{m!}}\frac{(r_{2}\beta)^{k}}{\sqrt{k!}}|t_{1}\alpha e^{-i(m+k)\theta}\rangle_{1}|m\rangle_{2}|r_{1}\alpha e^{-i(m+k)\theta}\rangle_{3}|k\rangle_{4}\nonumber\\
=\sum_{m,k=0}^{\infty}e^{-\frac{\alpha^{2}}{2}}\frac{(t_{1}\alpha)^{m}}{\sqrt{m!}}\frac{(r_{1}\alpha)^{k}}{\sqrt{k!}}|m\rangle_{1}|t_{2}\beta e^{-i(m+k)\theta}\rangle_{2}|k\rangle_{3}|r_{2}\beta e^{-i(m+k)\theta}\rangle_{4},
\label{loss}
\end{eqnarray}
where the second expression can be similarly obtained starting from the second identity in Eq. (1). One can see that the
entanglement between modes 1 and 2 leaks into the loss modes 3 and 4.
 
We quantify the entanglement in the presence of loss by numerically calculating the Logarithmic Negativity (LN) , which is defined as
$\log_{2}||\rho^{T_{A}}||_{1}$, where $\rho$ is the reduced density matrix for modes 1 and 2, and $\rho^{T_{A}}$ is its partial transpose. The LN is known to be an entanglement monotone \cite{key-1}. Note that the value of the LN for a two-qubit singlet state is 1. For simplicity we will focus on the case of symmetric loss, defining the loss percentage $\ell=r_1^2=r_2^2$.


\begin{figure}
\begin{centering}
\includegraphics[width=1\columnwidth]{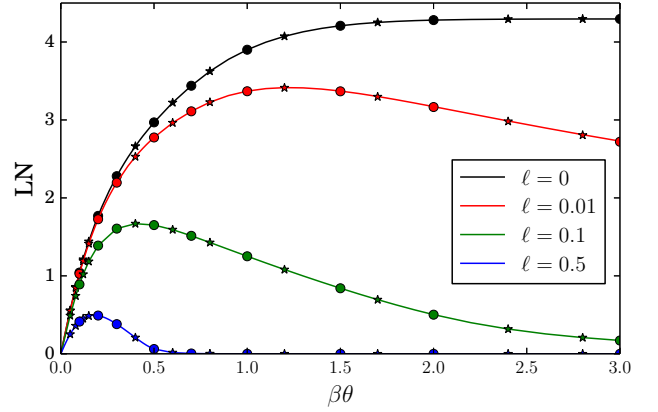}
\par\end{centering}
\centering{}\caption{Logarithmic negativity (LN) as a function of $\beta \theta$ for $\alpha=4$ and different values of $\beta$ and of the loss $\ell$. The solid curves are for $\beta=40$, with $\ell$ increasing from top to bottom. The results for $\beta=80$ (stars) and $\beta=160$ (circles) fall exactly on the curves for $\beta=40$, showing that the LN is only a function of the combination $\beta \theta$ in the considered regime ($\alpha \ll \beta, \theta \ll 1$). One sees that the entanglement saturates for $\beta \theta \gtrsim 1$ in the lossless case, but attains a maximum in the cases with loss.}
\end{figure}

We now discuss our results. Figure 2 shows three important points. First, as expected from our above discussion, the LN can take values that are significantly greater than one. Second, in the regime of large $\beta$ and small $\theta$ the LN is a function only of the product $\beta \theta$, which determines the separation between the neighboring coherent states as shown in Fig. 1 and in Eq. (1). One can see that the results for different values of $\beta$ and $\theta$ all fall on the same curve. This makes it possible to extrapolate our results to smaller values of $\theta$ and larger values of $\beta$, where a direct numerical calculation of the LN is not possible. For comparison, Ref. \cite{GaetaNP} demonstrated $\theta=0.0003$ with $\alpha=4$ and $\beta=450$, which gives $\beta \theta=0.135$. Fig. 2 shows that this would already be sufficient to demonstrate an LN of order one for reasonable loss values, with significantly increased LN possible for modestly improved $\beta \theta$.  Third, in the lossless case the entanglement saturates as a function of $\beta \theta$, exactly as expected from the discussion after Eq. (1). In contrast, in the presence of loss, there is an optimal value of $\beta \theta$. Greater separations between neighboring coherent states make the state more like a ``cat state'', and hence more fragile to photon loss. This is because for greater separations the states of the loss modes corresponding to the different terms in Eq. (2) become more orthogonal, leading to faster decoherence.

There is a similar phenomenon when $\alpha$ is increased, as shown in Figure 3. Fig. 3(a) shows the LN as a function of $\alpha$. One sees that in the lossless case, the entanglement increases monotonically with $\alpha$. With loss, this is no longer true. There is an optimal value for $\alpha$, which decreases as a function of the loss. This is further elucidated by Fig. 3(b), which shows the LN as a function of loss. One sees that the entanglement decreases with loss, which is not surprising. More interestingly, while greater values of $\alpha$ give greater entanglement for zero or very low loss, the entanglement decreases much faster for greater $\alpha$. This can be understood intuitively because greater values of $\alpha$ correspond to greater separations between the extremal terms in Eq. (1), see also Fig. 1. One can also view this result as a manifestation of a general trend that more macroscopic quantum effects tend to be more fragile \cite{wang}.

Figures 2 and 3 also show that there can be some entanglement even for large values of the loss (even more than 50 \%), provided that $\alpha, \beta$ and $\theta$ are chosen appropriately.

\begin{figure}
\begin{centering}
\includegraphics[width=1\columnwidth]{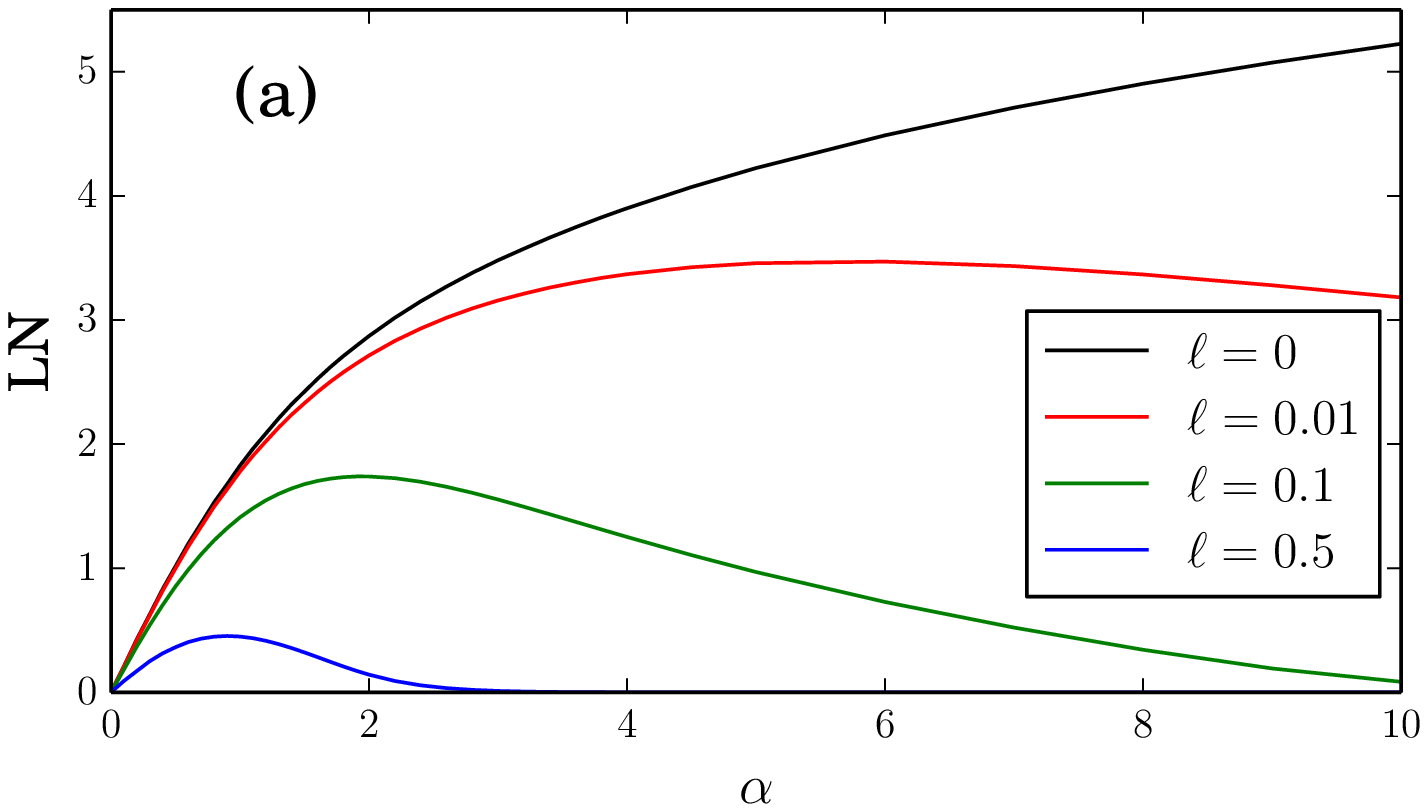}
\includegraphics[width=1\columnwidth]{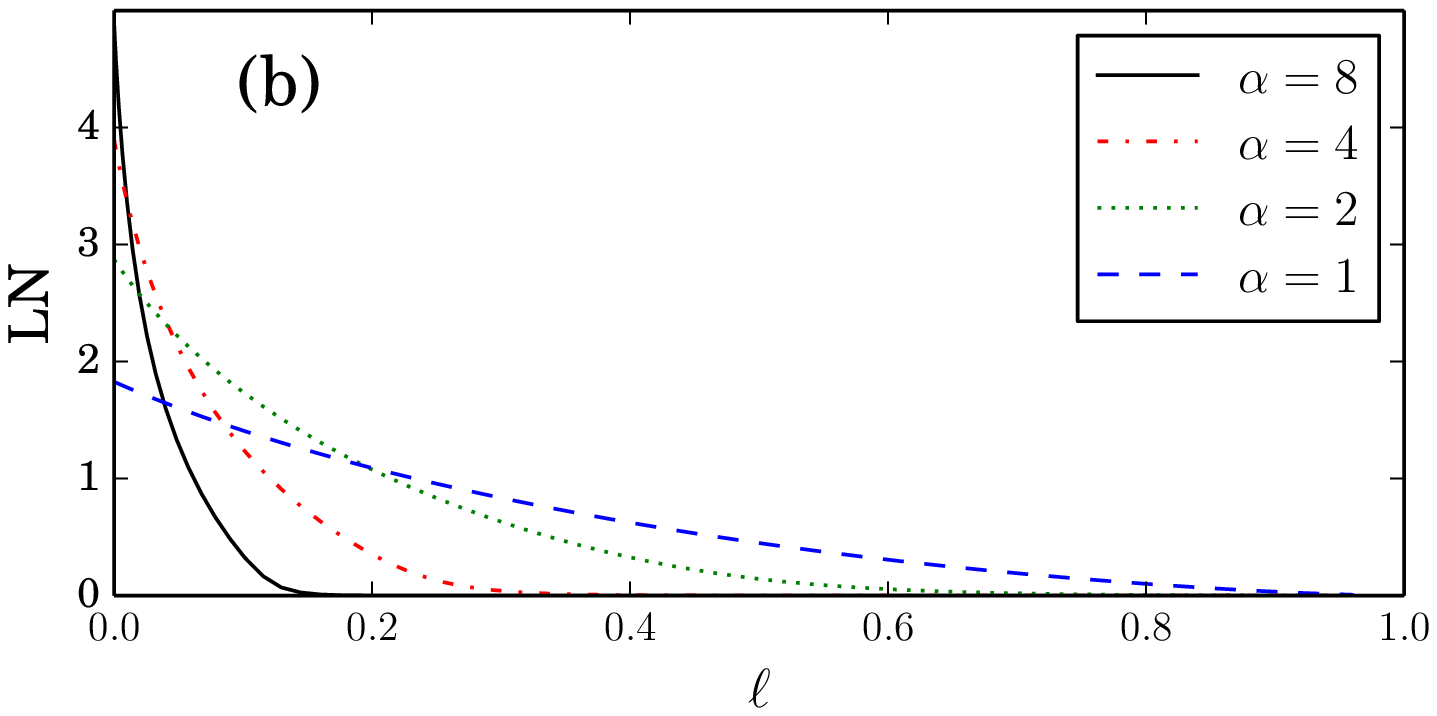}
\par\end{centering}
\centering{}\caption{(a) Logarithmic negativity (LN) as a function of $\alpha$ for different values of the loss $\ell$, for $\beta \theta=1$ ($\beta=50$). The entanglement increases with $\alpha$ in the lossless case, but attains a maximum in the presence of loss. (b) Logarithmic negativity as a function of loss $\ell$ for different values of $\alpha$, for $\beta \theta=1$ ($\beta=50$). Greater values of $\alpha$ yield greater entanglement in the lossless case, but this entanglement is more fragile under loss.}
\end{figure}

An experimental determination of the logarithmic negativity requires finding the density matrix, which can be done by homodyne tomography \cite{homtom}. Let us note that in the regime $\alpha \theta \ll 1$ that we are interested in here, the state of the macroscopic mode is well localized in phase space since it subtends an angle of order $2 \alpha \theta$ when viewed from the origin, see also Fig. 1. This means that it can be brought to the vicinity of the vacuum state by a phase space displacement, which should make tomography manageable even for large values of $\beta$. Nevertheless we now introduce an alternative approach, which does not require reconstruction of the full density matrix, but is based on an entanglement witness.

Our witness is inspired by the Duan criterion for two-mode squeezed states \cite{key-10}, which have EPR correlations of the form $\hat{x}_{1}+\hat{x}_{2}=0,\, \hat{p}_{2}-\hat{p}_{1}=0$. Analogously, our state has the number-phase correlations shown in Eq. (1), which can informally be written as $\hat{\phi}_{1}+\theta \hat{n}_{2}=0, \, \theta \hat{n}_{1}+\hat{\phi}_{2}=0$.
Of course the difficulty is that the phase operators $\hat{\phi}_1, \hat{\phi}_2$ are not well defined. Fortunately the quadrature operators $\hat{p}_1, \hat{p}_2$ can play an analogous role since
$\left\langle \alpha e^{-in\theta}|\hat{p}|\alpha e^{-in\theta}\right\rangle =\alpha\sin(n\theta) \approx \alpha \theta n$. Based on this intuition we define the operators
 \begin{equation}
\hat{u}=\hat{p}_{1}+\alpha \theta \hat{n}_{2}, \,
\hat{v}=\beta \theta \hat{n}_{1}+\hat{p}_{2}.
\end{equation}
Following closely the derivation of Ref. \cite{key-10} one can show that
for all separable states
\begin{align}
\left\langle \left(\Delta\widehat{u}\right)^{2}\right\rangle +\left\langle \left(\Delta\widehat{v}\right)^{2}\right\rangle  & \geq|\beta\theta\langle\hat{x}_{1}\rangle|+|\alpha\theta\langle\hat{x}_{2}\rangle|.
\end{align}
We can therefore define the entanglement witness
$w=\frac{|\beta\theta\langle\hat{x}_{1}\rangle|+|\alpha\theta\langle\hat{x}_{2}\rangle|}{(\Delta\hat{u})^{2}+(\Delta\hat{v})^{2}}$.
Whenever $w>1$ the state has to be entangled.

In practice it is advantageous to use a slightly modified witness which takes into account two important effects. First, under the time evolution $e^{-i\hat{n}_{1}\hat{n}_{2}\theta}$ each beam undergoes an average rotation, corresponding to a phase factor $e^{-i|\beta|^{2}\theta}$ for the first beam and $e^{-i|\alpha|^{2}\theta}$ for the second beam. This leads to very fast oscillations for the witness $w$. These oscillations can be eliminated by redefining the initial states as $\ensuremath{|\alpha\rangle_{1}=||\alpha|e^{i|\beta|^{2}\phi}\rangle_{1},\ensuremath{|\beta\rangle_{2}=||\beta|e^{i|\alpha|^{2}\phi}\rangle_{2}}}$, and by replacing the operators $u$ and $v$ defined above by $\hat{u'}=\hat{p}_{1}+\alpha(\hat{n}_{2}-|\beta|^{2})\theta,\hat{v'}=\hat{p}_{2}+\beta(\hat{n}_{1}-|\alpha|^{2})\theta$.  Second, in the presence of loss the amplitudes $\alpha$ and $\beta$ should be rescaled. This finally leads to the operators
\begin{equation}
\hat{U}=\hat{p}_{1}+t_{1}|\alpha|(\hat{n}_{2}-t_{2}^{2}|\beta|^{2})\theta \,,
\hat{V}=\hat{p}_{2}+t_{2}|\beta|(\hat{n}_{1}-t_{1}^{2}|\alpha|^{2})\theta
\end{equation}
and the final form of our witness
\begin{equation}
W=\frac{t_2 |\beta\theta\langle\hat{x}_{1}\rangle|+t_1|\alpha\theta\langle\hat{x}_{2}\rangle|}{(\Delta\hat{U})^{2}+(\Delta\hat{V})^{2}}\label{wit}. \end{equation}


\begin{figure}
\centering{}\includegraphics[width=0.7\columnwidth]{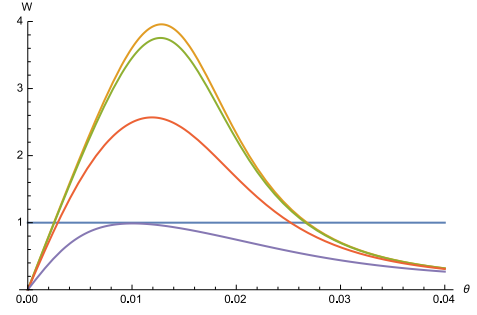} \caption{Entanglement witness $W$ of Eq. (\ref{wit}) as a function of $\theta$ for $\alpha=2, \beta=50$, and loss $\ell=0, 0.01, 0.1, 0.5$ from top to bottom. The witness shows entanglement above the line $W=1$.}
\end{figure}

Fig. 4 shows our results for the witness $W$. Most importantly, it allows one to detect entanglement for reasonable values of the loss, even though it is more sensitive to loss than the logarithmic negativity. In contrast to the latter, it attains a maximum as a function of $\theta$ even in the absence of loss, and the decay for $\theta$ increasing beyond that point is faster than for the logarithmic negativity. 

We have focused on the micro-macro regime, $\alpha \ll \beta$. While this is well motivated experimentally, we would still like to comment briefly on the macro-macro regime, $\alpha \sim \beta \gg 1$. The regime where $\alpha \theta \sim \beta \theta \ll 1$ is similar to two-mode squeezing. The clearly non-Gaussian regime where $\alpha \theta \sim \beta \theta \gtrsim 1$ (where there are many well-separated peaks in phase space) is  more difficult to analyze both theoretically and experimentally. Calculating the logarithmic negativity becomes intractable numerically, and our witness is also no longer suitable. Homodyne tomography also becomes much more difficult because the state can no longer be brought to the vicinity of the origin by simple displacements, see Fig. 1. However, our results in Fig. 3 suggest that the entanglement in this regime is very sensitive to loss, so experiments are likely to be unrealistic in any case for this reason.

We now discuss experimental demonstrations of the micro-macro entanglement analyzed above. We already discussed the fact that determining the logarithmic negativity requires homodyne tomography. The approach based on the witness $W$ does not require full reconstruction of the state, but it does require the ability to perform both homodyne detection (to measure $\hat{x}$ and $\hat{p}$) and photon counting (to measure $\hat{n}$). Loss causes ``vacuum noise'' contributions to the quadratures of order $\Delta x \sim \Delta p \sim \sqrt{\ell}$. It also causes an effective uncertainty in the photon number of order $\Delta n \sim \sqrt{\ell n}$, where $n$ is the total photon number, since the lost photons are essentially Poisson distributed. Our results on the effects of loss can therefore also be viewed as results on the necessary measurement precision, see also Ref. \cite{appel}. High-efficiency and low-noise homodyne detection was implemented e.g. in Ref. \cite{homodyne}, while photon counting of large photon numbers with a precision much better than $\sqrt{n}$ was demonstrated in Ref. \cite{beck}.

Some schemes for cross-phase modulation also generate self-phase modulation for one or both beams. However, the ladder scheme of Ref. \cite{GaetaNP}, for example, avoids self-phase modulation for the strong beam, whereas self-phase modulation for the weak beam is expected to be very weak when its amplitude is close to the single-photon level. There are also schemes that generate a pure cross-Kerr nonlinearity such as that of Ref. \cite{schmidt}. Note that self-phase modulation is a local operation that commutes with the cross-Kerr Hamiltonian, so it would not change the degree of entanglement (or the LN), but it would have to be taken into account or compensated when measuring the witness.

Our work was primarily motivated by recent experimental progress in the optical domain. Another promising system in this context is light stored in Bose-Einstein condensates, where significant cross-Kerr nonlinearities in combination with very low loss should be achievable \cite{BEC-kerr}. Large Kerr nonlinearities \cite{yale} and substantial cross-Kerr nonlinearities \cite{mw-cross} have already been obtained in the micro-wave domain using superconducting qubits. The present results could be of interest in these systems as well. Both in the optical and in the microwave domain it might be possible to map the created entangled state onto mechanical systems \cite{verhagen,lehnert}, which can be seen as further enhancing the degree of macroscopicity \cite{OMMM}.

As mentioned before, the situation considered in the present work can be seen as an amplification of the weak cross-Kerr nonlinearity by the strong coherent beam. We have seen that this can create strong micro-macro entanglement in a deterministic fashion. This approach to weak nonlinearities is thus distinct from the probabilistic weak-value based amplification scheme of Ref. \cite{feizpour} and the conditional entanglement generation schemes of Ref. \cite{conditional}. The non-Gaussian character of the generated entanglement could be beneficial for continuous-variable quantum information processing and quantum metrology \cite{oberthaler,non-gaussian}.

{\it Acknowledgments.} This work was supported by NSERC, AITF, WWTF and the Austrian Science Fund (FWF) through SFB FOQUS and the START grant Y 591-N16. We thank A. Feizpour, M. Girard, G. Gour, K. Heshami, A. Lvovsky, V. Narasimhachar and B. Sanders for useful discussions.

\end{document}